\documentclass[12pt,a4paper]{article}
\usepackage{hyperref}
\usepackage{authblk}
\usepackage{amsmath}
\usepackage{booktabs}
\usepackage{hyperref}
\usepackage{array}
\usepackage[T1]{fontenc}
\usepackage{amssymb}

\DeclareUnicodeCharacter{2212}{-}
\usepackage{times}

\title{\fontsize{18pt}{27pt}\selectfont
    { 
        HotStuff-2 vs. HotStuff: The Difference and Advantage}}
\author{\fontsize{12pt}{18pt}\selectfont
    {
        Siyuan Zhao,~Yanqi Wu,~Zheng Wang\thanks{Corresponding author}\\
            Shanghai Jiao Tong University \quad\\
            \small \texttt{\{zsy123,prphrl27,wzheng\}@sjtu.edu.cn} \\
    }
 }
\date{}
\usepackage{amsmath,amsfonts,amssymb}
\usepackage{graphicx}
\usepackage{subfigure}
\usepackage{float}
\usepackage[export]{adjustbox}
\usepackage{bibentry}
\usepackage{abstract}

\usepackage{xcolor}

\usepackage{url}
\usepackage{bbding}
\usepackage{bm}
\usepackage{longtable}
\usepackage{supertabular}
\usepackage{changepage}
\usepackage{enumerate}
\usepackage{caption}
\usepackage{indentfirst}
\usepackage[left=2.50cm,right=2.50cm,top=2.80cm,bottom=2.50cm]{geometry}

\usepackage{fancyhdr} 
\hypersetup{colorlinks=true,linkcolor=black}

\begin{document}
    \maketitle
    \renewcommand{\refname}{References}

    \lhead{}
    \chead{}
    \rhead{}
    \lfoot{}
    \cfoot{\thepage}
    \rfoot{}
   \begin{center}
        \textbf{Abstract}
    \end{center}

    \noindent
    \indent Byzantine consensus protocols are essential in blockchain technology. The widely recognized HotStuff protocol uses cryptographic measures for efficient view changes and reduced communication complexity. Recently, the main authors of HotStuff introduced an advanced iteration named HotStuff-2. This paper aims to compare the principles and analyze the effectiveness of both protocols, hoping to depict their key differences and assess the potential enhancements offered by HotStuff-2.

    \vspace{2ex} 
    \vspace{2ex} 

\section{Introduction: HotStuff and HotStuff-2}
\subsection{The HotStuff Consensus Protocol}
\begin{figure}[!tb]
	\centering
	\begin{minipage}{0.83\textwidth}
		\centering
		\includegraphics[width=1.0\textwidth]{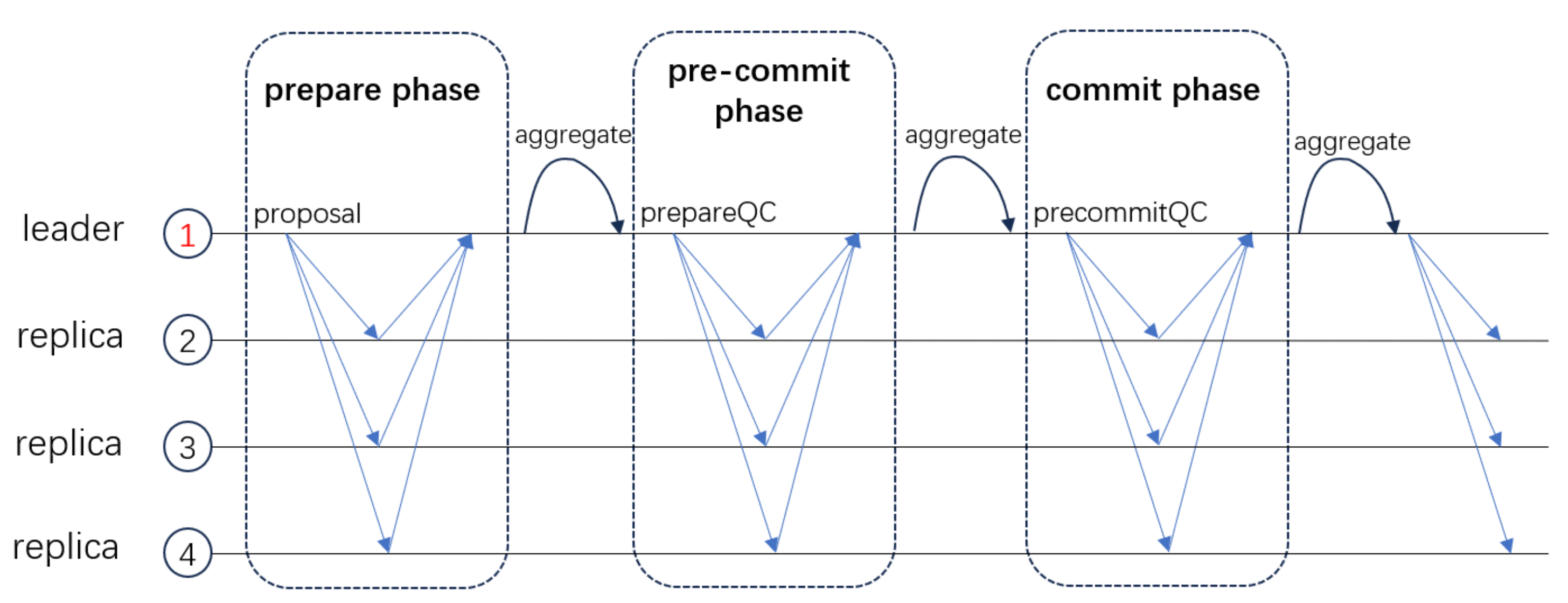}
		\caption{\fontsize{10pt}{15pt}\selectfont The Consensus Process of HotStuff}
             \label{HotStuff}
	\end{minipage}
\end{figure}
The consensus process of HotStuff~\cite{yin2019hotstuff} protocol is illustrated in Figure~\ref{HotStuff}, where the protocol segregates the entire consensus mechanism into four distinct phases: prepare, pre-commit, commit, and decide.  In a network comprising $n$ consensus nodes and $f$ Byzantine nodes, with the condition that $n\geq3f+1$, the operational procedure for each phase in HotStuff adheres to a consistent pattern. The leader node aggregates $n−f$ votes from the preceding round to create a Quorum Certificate (QC), then broadcasts this QC. Subsequently, each consensus node casts a vote on the received QC and forwards it to the leader node. The HotStuff consensus protocol ensures liveness through its Pacemaker mechanism, which facilitates automatic view changes via a timeout strategy, thus preventing any liveness impairment. Meanwhile, the safety of HotStuff is secured by its SafeNode rule. 
When compared with the classical PBFT~\cite{castro1999practical} protocol, HotStuff's principal enhancements are as follows:
\paragraph{1. Linear and Effortless View Changes.} HotStuff distinguishes itself with a streamlined communication approach, where the interaction is primarily between the consensus nodes and the leader, rather than the all-to-all broadcast paradigm employed in PBFT. This modification curtails the communication complexity from $O(n^{2})$, as seen in PBFT, to a generally linear scale. Additionally, HotStuff ingeniously treats messages pertaining to new-view changes as regular consensus messages. This means that, upon collecting $n-f$ new-view messages, the leader will discontinue the current consensus and initiate the prepare phase of the subsequent round. This approach avoids the extra cost in PBFT, where each consensus node must send a new-view message containing $n − f$ messages from the previous round to the leader, generating significant overhead.
\paragraph{2. Optimistic Responsiveness.} Based on a partially synchronous network model, HotStuff incorporates a feature known as optimistic responsiveness. This characteristic implies that after Global Stabilization Time (GST)~\cite{dwork1988consensus}, any non-Byzantine leader node can effectively drive the consensus process forward simply by gathering $n-f$ votes, without necessitating additional delays. This property is consistently maintained even in case of view changes. Such optimistic responsiveness sharply contrasts with the approach taken by the Tendermint~\cite{buchman2016tendermint} consensus protocol. In Tendermint, there is a compulsory waiting delay following a view change, which is implemented to safeguard the protocol's safety. However, this will influence the consensus process, because the protocol must pause for this fixed duration before proceeding. This makes HotStuff more effective during view change, without affecting its security.
    
\subsection{The HotStuff-2 Consensus Protocol}
\begin{figure}[!tb]
	\centering
	\begin{minipage}{0.9\textwidth}
		\centering
		\includegraphics[width=1.0\textwidth]{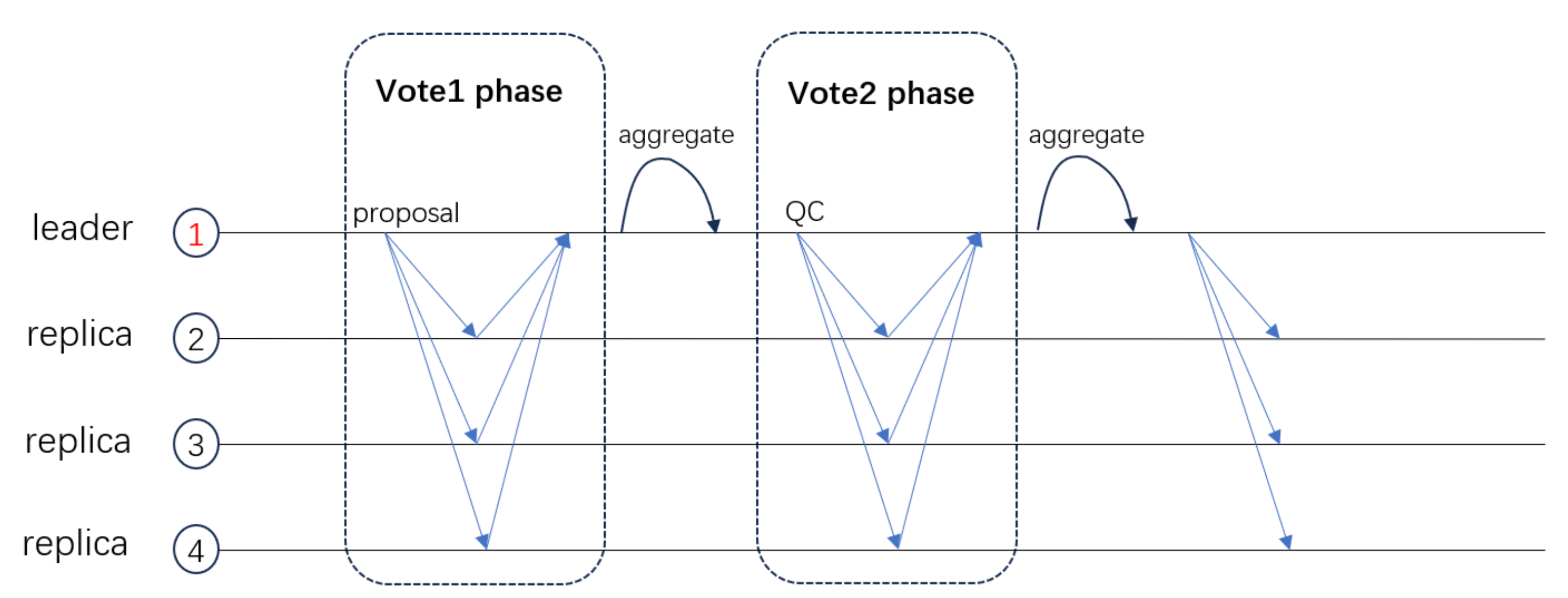} 
		\caption{\fontsize{10pt}{15pt}\selectfont The Consensus Process of HotStuff-2}
            \label{hotstuff-2}
	\end{minipage}
\end{figure}
HotStuff-2~\cite{malkhi2023HotStuff}, an advancement of the original HotStuff protocol, is visualized in Figure~\ref{hotstuff-2}. This method streamlines the consensus process by reducing one voting round, thus cutting down cryptographic cost while preserving linear view changes and optimistic responsiveness. Introducing the happy-path concept, HotStuff-2 employs two-phase voting under favorable conditions with non-Byzantine sequential leader nodes. This innovation ensures efficiency and security across various network scenarios, optimizing performance in ideal conditions.

When faced with deteriorating network conditions or the emergence of Byzantine leader nodes, HotStuff-2 safeguards its safety and liveness via the Pacemaker mechanism. In these situations, frequent view change is common, leading to differences in view height of nodes. To address this, the Pacemaker activates a view synchronization waiting mechanism for all non-leader nodes that are lagging. This mechanism requires these nodes to pause and wait for the leader to broadcast the proposal carrying the latest view. Once received, the nodes update their local state and subsequently reboot the two-phase voting consensus process. This ensures that all nodes remain synchronized and actively participate in the consensus, preserving the security and liveness of the protocol.

In summary, HotStuff-2 demonstrates adaptive behavior contingent on network conditions and the characteristics of the leader nodes. It strategically incorporates a $\delta$ time delay in scenarios with malicious leaders to ensure protocol security. During happy-path states, characterized by favorable conditions and non-Byzantine leaders, HotStuff-2 efficiently executes two-phase voting. This nuanced approach greatly boosts HotStuff-2's performance, marking a substantial optimization compared to the original HotStuff protocol.

\begin{figure}[!t]
	\centering
	\begin{minipage}{0.63\textwidth}
		\centering
		\includegraphics[width=1.3\textwidth,center]{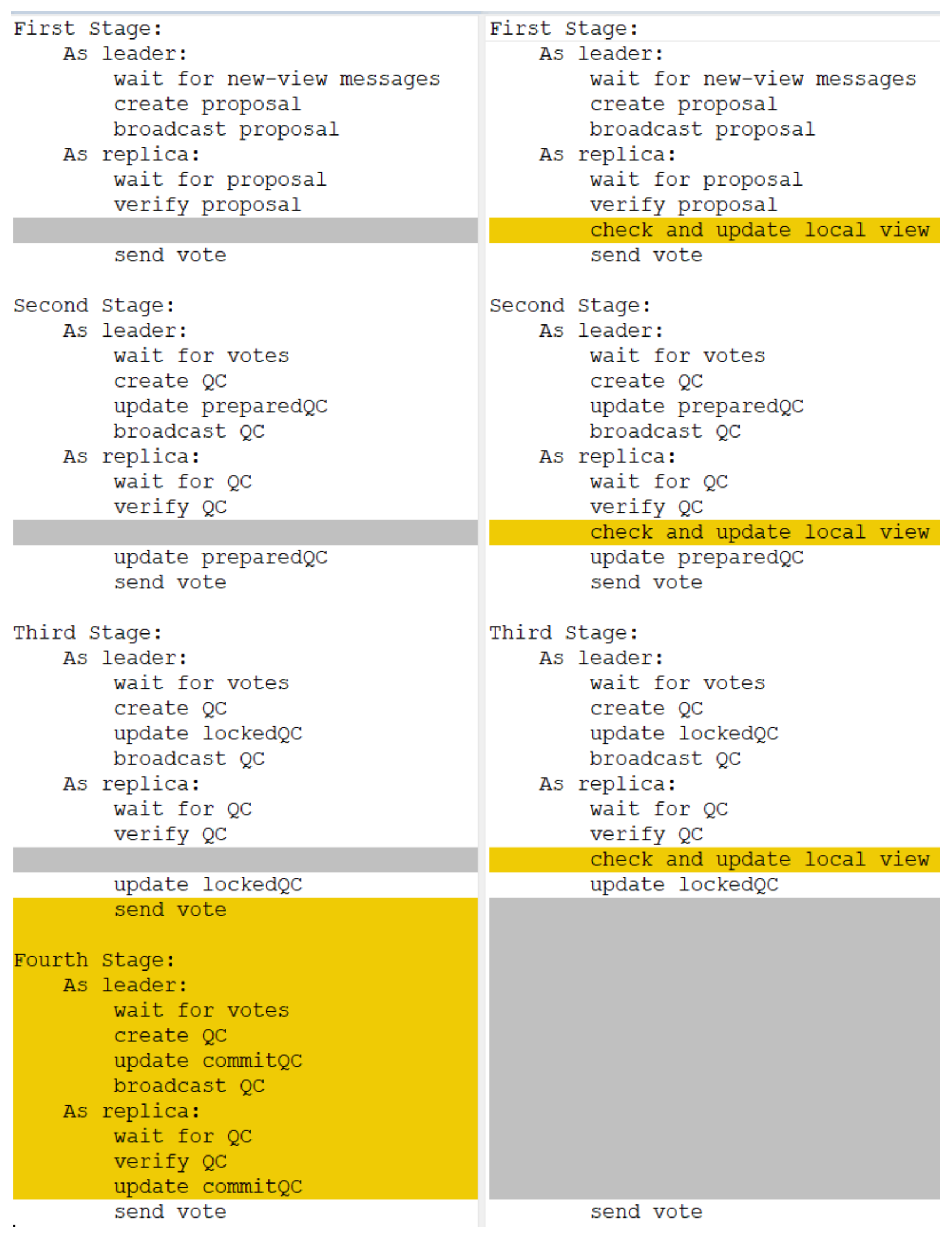}
		\caption{\fontsize{10pt}{15pt}\selectfont Pseudo-code Comparison between HotStuff and HotStuff-2}
        \label{code}
	\end{minipage}
\end{figure}

\section{Comparing HotStuff and HotStuff-2}
Figure~\ref{code} provides a pseudo-code comparison between the HotStuff and HotStuff-2 protocols, offering a visual representation of the variations in code volume between them. This comparison effectively illustrates the streamlined approach of HotStuff-2. Furthermore, Table~\ref{parameter} presents a comparison of the consensus parameters for both protocols, highlighting various factors that contribute to the efficiency of the consensus process. Overall, HotStuff-2, in its evolution from the original HotStuff, incorporates key modifications, which primarily include:
\begin{enumerate}
  \item \textbf{HotStuff-2 simplifies the consensus process.} HotStuff-2 optimizes the consensus process by reducing one voting phase, enhancing efficiency compared to the original HotStuff's three rounds.
  \item \textbf{HotStuff-2 introduces view synchronization waiting mechanism.} This is activated when nodes are not in sync, employing a timeout feature to align their views, thereby ensuring cohesive progression in the consensus process.
\end{enumerate}

\begin{table*}[!t]
\caption{Consensus Parameters Comparison between HotStuff and HotStuff-2}
\centering
\small 
\begin{tabular}{
    >{\raggedright\arraybackslash}p{2.5cm} 
    >{\raggedright\arraybackslash}p{2.5cm} 
    >{\raggedright\arraybackslash}p{2.5cm} 
    >{\raggedright\arraybackslash}p{2.5cm} 
    >{\raggedright\arraybackslash}p{2.5cm} 
}
\toprule
Consensus Parameter &  Consensus Nodes &  Byzantine Nodes &   Communication Delay &  View Switch Delay \\
\midrule
 HotStuff   & \CheckmarkBold & \CheckmarkBold & \CheckmarkBold & ~ \\
 HotStuff-2 & \CheckmarkBold & \CheckmarkBold & \CheckmarkBold & \CheckmarkBold \\
\bottomrule
\end{tabular}
\label{parameter}
\end{table*}

Due to these modifications, theoretically, HotStuff-2 holds two potential advantages over the HotStuff protocol: 1) HotStuff-2 is simpler and easier to implement; and 2) In scenarios where view synchronization isn't necessary, HotStuff-2 exhibits superior operational efficiency.

\section{Experiments}
To investigate the distinctions in design and performance between HotStuff and HotStuff-2, we execute a series of experimental analyses. These experiments are conducted using Python 3.9.7 to implement both protocols, on a system equipped with Windows 11. In the experimental setup, we simulate multiple consensus nodes within a single process, utilizing a hash table to simulate the public key signing and verification process. In addition, we integrate a basic communication delay function to simulate the communication delay between nodes. We assess the efficiency of each consensus protocol by measuring the process run time over a consistent number of consensus rounds, specifically across 10 iterations. This methodical approach allows for a comparative analysis of the two protocols under controlled and replicable conditions, highlighting their respective efficiencies and operational differences.

\subsection{Influence of Communication Delay}
Initially, we fix the consensus node number $n=13$ and the Byzantine node number $f=4$, satisfying the $n\geq3f+1$ condition required for BFT consensus protocols. Moreover, we set the view switch delay for HotStuff-2 at 0.5 seconds and communication delays ranging from 10 to 0.0001 seconds. 

\begin{table*}[!tb]
\caption{Influence of Communication Delay on Efficiency}
\centering
\begin{tabular}{ccccccc}
\toprule
 Communication Delay & 10s & 1s & 0.1s & 0.01s & 0.001s &0.0001s  \\ 
\midrule
       
        HotStuff & 400.37s & 40.49s & 4.33s & 0.62s & 0.61s & 0.59s  \\
        HotStuff-2 & $\bm{300.25s}$ & $\bm{30.34s}$ & $\bm{3.26s}$ & $\bm{0.47s}$ & $\bm{0.45s}$ & $\bm{0.43s}$  \\
\bottomrule
\end{tabular}
\label{delay}
\end{table*}

As shown in Table~\ref{delay}, HotStuff-2 consistently surpasses HotStuff in efficiency across various communication delays. This may due to HotStuff-2's fewer voting rounds, which reduce communication and cryptographic cost, especially in scenarios with fewer Byzantine nodes.

\begin{table*}[!tb]
\caption{Influence of Consensus Node Numbers on Efficiency}
\centering
\scalebox{0.8}{
\begin{tabular}{cccccccccccc}
\toprule
Consensus Node & 13 & 22 & 31 & 40 & 49 & 58 & 67 & 76 & 85 & 94 & 103  \\
\midrule
       
        HotStuff & 4.33s & 4.38s & 4.36s & 4.37s & 4.37s & 4.38s & 4.38s & 4.40s & 4.42s & 4.43s & 4.44s\\ 
        HotStuff-2 & $\bm{3.25s}$ & $\bm{3.28s}$ & $\bm{3.28s}$ & $\bm{3.28s}$ & $\bm{3.29s}$ & $\bm{3.29s}$ & $\bm{3.29s}$ & $\bm{3.28s}$ & $\bm{3.30s}$ & $\bm{3.31s}$ & $\bm{3.31s}$  \\ 
\bottomrule
\end{tabular}}
\label{nodes}
\end{table*}

\subsection{Influence of Consensus Node Number}
In this experiment, we configure the consensus nodes number $n$ increasing from 13 to 103, in intervals of 9. Correspondingly, we maintain the number of Byzantine nodes at one-fourth of the total nodes, ensuring adherence to the $n\geq3f + 1$ consensus condition. Furthermore, we establish the communication delay at 0.1 seconds and set the view switch delay specific to HotStuff-2 at 0.5 seconds. 

As shown in Table~\ref{nodes}, HotStuff-2 consistently surpasses HotStuff in efficiency, with increasing the consensus node number. This trend emphasizes the enhanced scalability of HotStuff-2 due to streamlined design and protocol optimizations, improving performance in larger network environments.

\subsection{Influence of Byzantine Node Numbers}
Initially, we fix the consensus nodes number $n=103$ and set the Byzantine node number $f$ increasing from 0 to 34 in increments of 5, thereby always satisfying the $n\geq3f+1$ consensus condition. Subsequently, we set the communication delay to 0.1 seconds and the HotStuff-2 view change delay to 0.5 seconds. 

\begin{table*}[!tb]
\caption{Influence of Byzantine Node Numbers on Efficiency ($n=103$)}
\centering
\scalebox{1.0}{
\begin{tabular}{cccccccc}
\toprule
Byzantine Node & 4 & 9 & 14 & 19 & 24 & 29 & 34 \\
\midrule
       
        HotStuff   & 4.42s        & 4.43s        & 4.43s        & 4.45s        & $\bm{4.45s}$ & $\bm{4.46s}$  & $\bm{4.48s}$ \\
        HotStuff-2 & $\bm{3.45s}$ & $\bm{3.63s}$ & $\bm{4.01s}$ & $\bm{4.39s}$ & 4.85s        & 5.21s         & 6.03s       \\
\bottomrule
\end{tabular}}
\label{bnodes}
\end{table*}

\begin{figure}[!tb]
    \centering
    \begin{minipage}{0.83\textwidth}
        \centering
        \includegraphics[width=1.0\textwidth]{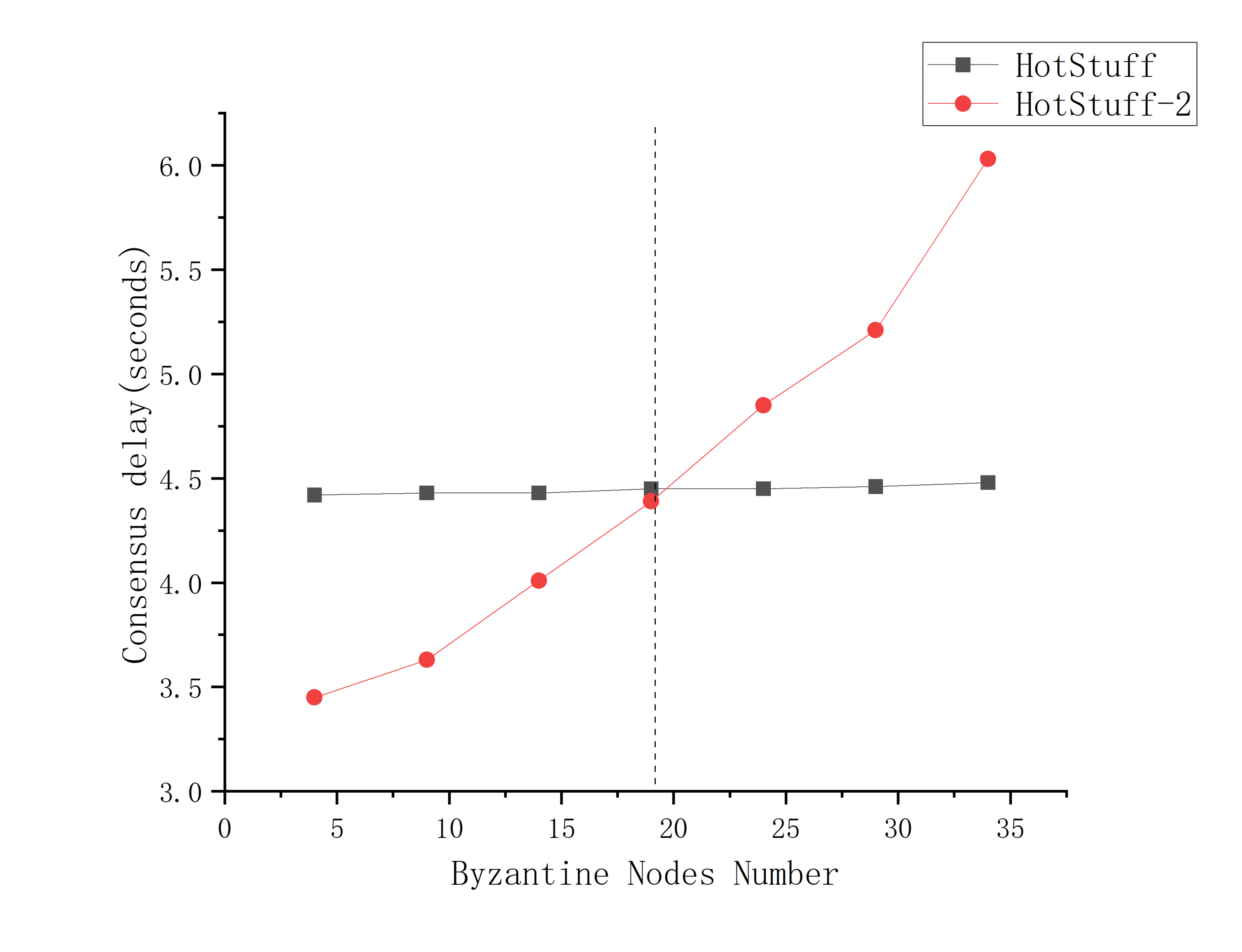} 
        \caption{\fontsize{10pt}{15pt}\selectfont Comparison of Consensus Efficiency Under Different Numbers of Byzantine Nodes}
        \label{byzantie}
    \end{minipage}
\end{figure}

As shown in Table~\ref{bnodes} and Figure~\ref{byzantie}, HotStuff-2 is more effective than HotStuff when the number of Byzantine nodes is low. However, as the number of Byzantine nodes approaches one-fifth of the total nodes, HotStuff shows higher efficiency. This suggests that the efficiency of HotStuff-2 is contingent upon the rate of Byzantine nodes in the network, as it is more stable and faster in scenarios with lower rate of Byzantine nodes. Conversely, with a larger rate of Byzantine nodes, the frequent need for view change incurs considerable additional cost.

\section{Conclusion}
This study presents a comparative analysis of two well-known Byzantine consensus protocol: HotStuff, and its enhanced iteration, HotStuff-2. To facilitate this comparison, we implement both protocols within the same framework and conduct a series of experimental evaluations. Our findings highlight two primary advantages of HotStuff-2 over the original HotStuff: 1) HotStuff-2 exhibits a more compact code, which translates to ease of implementation; and 2) HotStuff-2 shows superior operational efficiency, particularly in network environments with a lower rate of Byzantine nodes. These advantages underscore HotStuff-2's advancements in both usability and performance within the realm of Byzantine consensus protocols.

\bibliographystyle{unsrt}
\bibliography{main}

\end{document}